# Fine structure of histograms of alpha-activity measurements depends on direction of alpha particles flow and the Earth rotation: experiments with collimators.

**Simon E. Shnoll**[*†], **Konstantin I. Zenchenko**[†], **Iosas I. Berulis**[‡], **Natalia V. Udaltsova**[*] and **Ilia A. Rubinstein**[§]

[*] *Institute of Theoretical and Experimental Biophysics of RAS, Pushchino, Moscow Region, 142290, Russian Federation*

[†] *Department of Physics, Moscow State University, Vorob'evy Gory, Moscow, 119992, Russian Federation*

[‡] *Radioastronomical Station, Pushchino, 142290, Russian Federation*

[§] *Skobeltsyn Institute of Nuclear Physics, Moscow State University, Moscow, 119992, Russian Federation*

Shnoll@iteb.ru, udanat@yahoo.com

**Summary**

The fine structure of histograms of measurements of $^{239}$Pu alpha-activity varies periodically, and the period of these variations is equal to sidereal day (1436 minutes). The periodicity is not observed in the experiments with collimator that restricts the alpha particles flow to be oriented to the Polar Star. Based on this study and other independent data, such as measurements conducted by the Arctic expedition, and similarity of the histograms in processes observed at different locations at the same local time, the conclusion was made, that the fine structure of statistical distributions of the observed processes depends on the celestial sphere.

**Introduction**

During many years, we explore changes in the fine structure of the histograms for consequent measurements of parameters of various physical processes. We call such changes as "macroscopic fluctuations" (**MF**). (Udaltsova, N.V. et all. 1987; Shnoll S.E. et all.1992, 1998, 2000; Fedorov M.V. et all.2003).

Comparing various physical processes (from measurements of noise in gravitation-gradient antenna and rates of biochemical reactions, to radioactive decay measurements) observed at different geographic locations at the same local (longitudinal) time, we have found the following:

1. Shapes of histograms in such different processes are similar at the same local time with high likelihood ( Shnoll S.E.2001).

2. The shape of histograms varies regularly in time. Here are some observations:

a. The histograms of the adjacent time intervals are similar with highest likelihood: this is the "neighbor zone effect".

b. For different locations similar histograms are observed most likely at the same local time: this fact indicates that histogram shape depends on the rotation of the Earth around its axis.

c. The recurrence of the similar shape of the histogram is observed with the period 1436 minutes, i.e. sidereal day.

These effects can be explained by the fact that, while the Earth is rotating around its axis and moves at the same time around the Sun, different sections of the Earth's surface are exposed to different sectors of the celestial sphere (Shnoll S.E., 2001; Fedorov et all. 2003).

The assumption that the shape of histograms depends on the celestial sphere "above the place of measurements", leads to the next ones: "the neighbour zone effect" and the about-daily variations of the histogram shape disappear at the geographical poles.

This hypothesis was confirmed in the study of $^{239}$Pu alpha-activity during the Arctic expedition at the Arctic Ocean. This expedition took place in August – October, 2000; the measurements were conducted at 82 degree of the North latitude, aboard the ship "The Academician Fedorov" ( Shnoll S.E.et all., 2003).

Additional confirmations of this assumption were obtained in the study of collimators restricting alpha particle flow in certain directions. The results of these experiments are presented in this article.

**Objects and methods**

The most convenient object for the MF studies is the process of radioactive decay, especially alpha decay. Alpha decay is definitely immune to any trivial factors. It is possible to register separate acts of radioactive decay as a "binary" phenomenon (in logic of 0, 1).

This possibility was exercised in the specially constructed sensitive devices designed by one of the authors (I.A.R.). These devices provide transformation of the registered alpha activity data into time-series datasets with absent low-frequency trends. The results of the measurements according to all criteria correspond to "white noise" with Poisson distribution.

We used two identical devices for alpha-radioactivity measurements of $^{239}$Pu samples. Both devices provided rigid fixation of the source position in respect to the detector. The collimator (which "cuts out" a jet of alpha particles in a specified direction) may be placed between the source and the detector. One of the devices had the collimator oriented toward the Polar Star. The other device in one case had no collimator and the detector registering alpha particles emitting from the entire surface of the source which was West-oriented, in other case it had the collimator oriented to the West.

The distance between the source and the detector was 12 mm for both devices. While flying this distance, alpha particles lose about 10% of their energy. As the result, the energy of particles reaching the detector was ~ 4 MeV. The registration threshold was ~ 1,6 MeV; it completely excluded any possible influence of detector noise as well as humidity and air density variations.

A crystal generator (131 MHz frequency) was used for measuring time intervals. The power voltage of the converter was stabilized. The instability of the registration threshold was about ± 6% for the temperature range from minus $30^0$ C to plus $50^0$ C.

**Fig. 1** shows the histogram constructed for 60 of the results of consecutive measurements of alpha-activity of the $^{239}$Pu sample. The duration of 1 measurement is 1 second. X-axis represents the measured parameter (in this case, the number of registered events of alpha decay). Y-axis shows the number of measurements in respect to the observed level of activity.

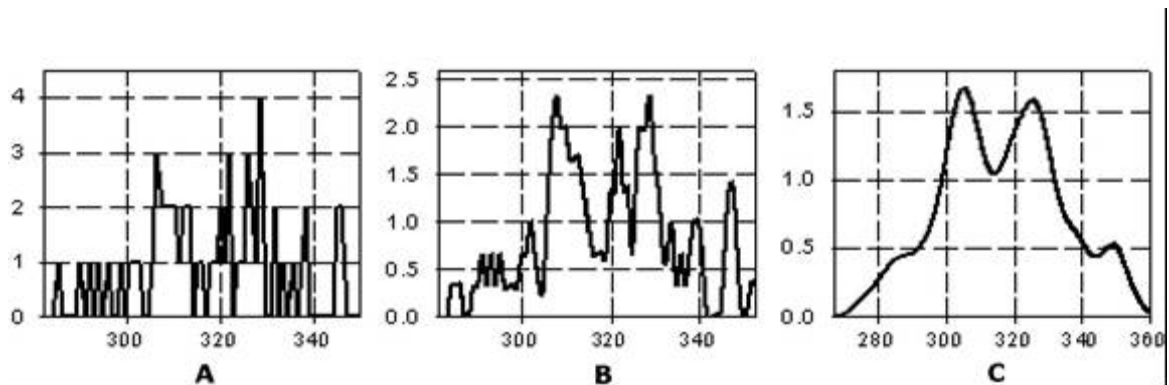

**Fig. 1**
**A. Non-smoothed "inconsistent" histogram, 60 measurements.**
**B. The same histogram, after single smoothing.**
**C. The same histogram smoothed 15 times.**

As shown the actual value of the ordinate is quite small, not more than 4. The number of classes (in this case, it is 60) used for constructing these histograms is compatible to the number of measurements (in this case, it is also 60). As a result the original histogram is pockmarked. Such histograms are called "inconsistent". It is hard to find any regularity in change of shape in such histograms. However, things alter when we smooth these insufficient histograms using the method of moving averages. Look at **Fig**. 1-**B**; it is the same histogram after single smoothing. The result of 15 times smoothing is shown at **Fig. 1-C**. It is evident that, as a result of smoothing, the specific shape becomes visible. In the following discussion we consider just these histograms

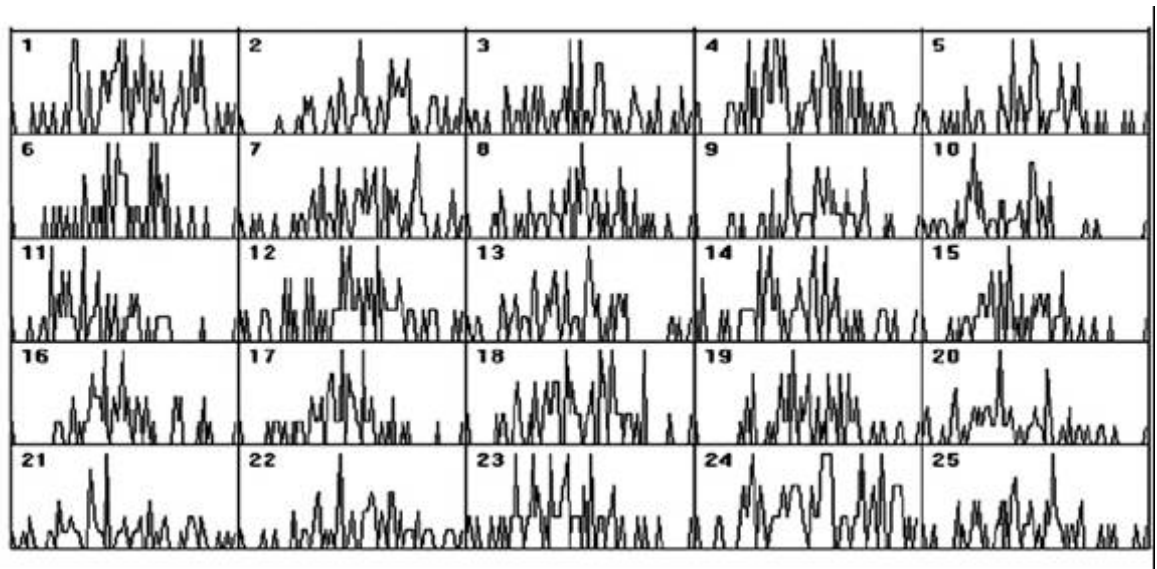

Fig. 2. Non-smoothed histograms constructed for non-overlapped sections of time-series ; every histogram represents 60 results of alpha-activity measurements in the $^{239}$Pu sample, the time interval for each histogram is 1 minute. Axes are the same as at Fig. 1.

.

**Fig. 2** represents a number of non-smoothed inconsistent histograms constructed for the first 25 sections of time-series results of alpha-activity measurements in the $^{239}$Pu sample. (these sections are not overlapping). Each section covers 60 measurements within 1 minute in total. One couldn't find evident regularities for these distributions.

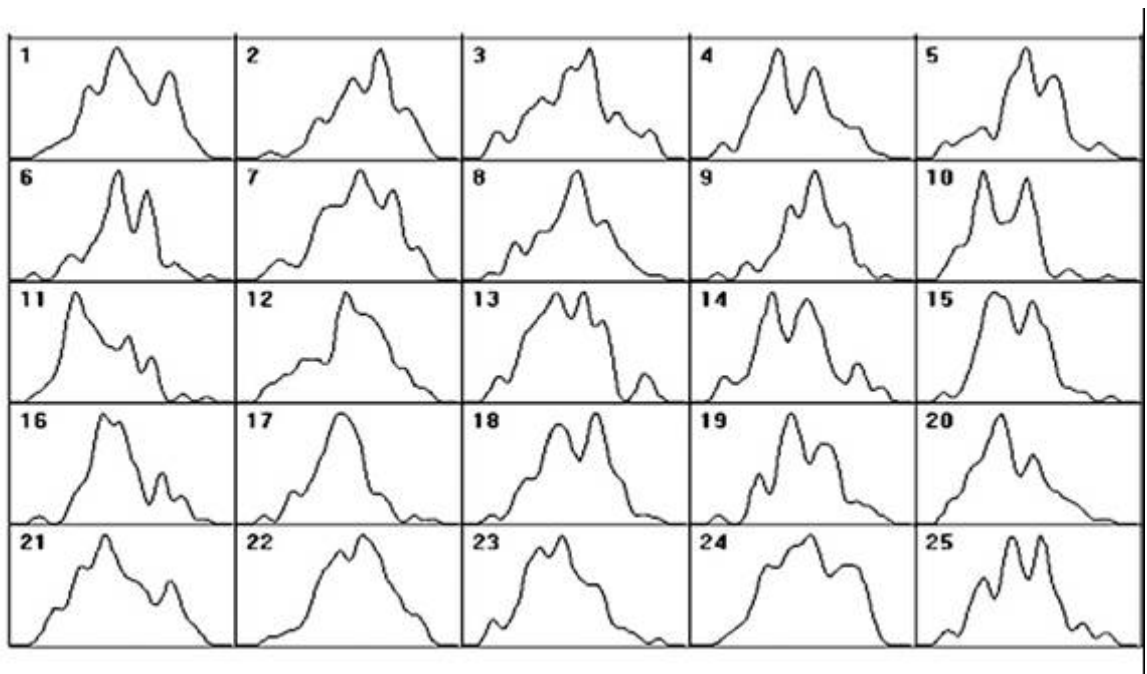

Fig.3. The same histograms as at Fig. 2 smoothed 15 times. Axes are the same as at Fig. 2.

Fig. 3 represents the same histograms as at **Fig. 2**, after 15 times smoothing. The similar form of some histograms becomes evident. Is such a similarity accidental? To answer this question, we have to compare thousands of possible combinations of histogram pairs.

For this purpose, we used the computer program "Histogram Manager" (HM) designed by Edwin Pozharsky (Shnoll S.E. et all. 1998a). This program provides construction of histograms, their smoothing, imposition upon each other, some linear transformations, as well as the accumulation of the records of all selected similar pairs in a special file. The main function of this program is the calculation and construction of the frequency distribution of selected similar histograms against time-intervals between them. Nevertheless an expert performs the most important part of the analysis - the diagnosis "similar – not similar".

The comparison of many thousands of histogram pairs is a job extremely time and labor consuming. There have been many attempts to replace an expert with an adequate computer program, but all the attempts failed. The various methods were used - from regular calculations of correlation coefficients, statistical criteria estimation, to application of neural networks and Wavelet-analysis. The results of application of all these methods are still inferior to a visual estimation.

Some efforts were made to eliminate the bias in the diagnostic process. The program HM does this through the randomization of histogram numbers. In this case, the expert is blinded to the histogram numbers; so the distribution of the time-intervals between the similar histograms obtained by the expert may not be biased.

The main purpose of the study was investigation of the association between the realization of similar histograms and the value of time interval between them. Based on the results of histogram comparison, we construct the distribution of a number of similar pairs in respect to the length of the interval between them.

Many thousands of histogram pairs were compared to get the reliable distributions for similar pairs of histograms in respect to time intervals.

The estimation of statistical significance of maximums of these distributions was based on hypergeometric distribution (Pearson E.S. et all. 1958.) As shown in the work cited (N.V.Udaltsova see in: Shnoll S.E. et all. 2004), the Poisson distribution may be used as a general, broader estimation of the significance. We will utilize this method for conclusions that have a fundamental value.

**Results**

From January 31 through July 11, 2002, and from February 19 through December 2003, ten sets of round-o-clock measurements of $^{239}$Pu alpha-activity were conducted at the Institute of Theoretical and Experimental Biophysics, RAN (RUSSIAN ACADEMY OF SCIENCE), Pushchino (latitude: $54^0$ 50' North; longitude: $37^0$ 38' East). Measurements were taken every second using two independent identical devices. One of the devices was equipped with the collimator oriented toward the Pole Star, the other one was either equipped with a collimator oriented to the West, or without collimator with detector and a radioactive sample plate oriented to the West. Each set of measurements continued about one month. The results of one-second measurements were summarized into one-minute rates. Every 60 one-minute numbers were combined into a histogram, so each histogram was built on one-hour interval of time-series. As a result we "replaced" a set of radioactive decay measurements by a set of consecutive one-hour histograms. After that we made a one-by-one comparison of smoothed histogram shapes and constructed a frequency chart of similar histograms against the time interval between them.

The results of experiments of 2002 – 2003 are presented in the table 1 and at the **fig 4.** In the **Table 1** counts of similar histograms are presented against the time interval between them. Measurements were conducted with no collimator (February 17, 2002 – July 11, 2002), and with collimator oriented to the West (February 19, 2003 – December 2, 2003).

**Table 1.**

**Counts of similar histograms of measurements of $^{239}$Pu alpha-activity by time interval between histograms (a flat detector without the collimator was used).**

| Interval (hours) | 17 Feb 2002 | 18 Mar 2002 | 15 Apr 2002 | 14 Jun 2002 | 11 Jul 2002 | 19 Feb 2003 | 19 Feb 2003 a | 18 Mar 2003 | 15 Apr 2003 | 2 Dec 2003 | Total |
|---|---|---|---|---|---|---|---|---|---|---|---|
| 1 | 118 | 79 | 86 | 66 | 76 | 79 | 71 | 89 | 93 | 133 | 890 |
| 2 | 65 | 36 | 44 | 35 | 64 | 42 | 34 | 47 | 38 | 78 | 483 |
| 3 | 76 | 30 | 39 | 36 | 50 | 34 | 31 | 25 | 31 | 48 | 400 |
| 4 | 60 | 38 | 53 | 34 | 27 | 31 | 31 | 34 | 23 | 41 | 372 |
| 5 | 45 | 40 | 28 | 31 | 40 | 36 | 38 | 55 | 39 | 75 | 427 |
| 6 | 27 | 32 | 20 | 21 | 42 | 28 | 24 | 37 | 26 | 68 | 325 |
| 7 | 34 | 48 | 29 | 17 | 46 | 33 | 32 | 36 | 27 | 63 | 365 |
| 8 | 45 | 30 | 29 | 22 | 24 | 38 | 36 | 38 | 18 | 67 | 347 |
| 9 | 37 | 32 | 27 | 20 | 31 | 31 | 28 | 34 | 21 | 68 | 329 |
| 10 | 41 | 36 | 29 | 35 | 33 | 44 | 25 | 47 | 18 | 55 | 363 |
| 11 | 51 | 44 | 57 | 30 | 49 | 47 | 34 | 32 | 19 | 74 | 437 |
| 12 | 37 | 34 | 42 | 27 | 45 | 44 | 38 | 26 | 28 | 64 | 385 |
| 13 | 34 | 36 | 39 | 23 | 40 | 43 | 34 | 31 | 17 | 67 | 364 |
| 14 | 52 | 22 | 38 | 20 | 50 | 52 | 43 | 34 | 39 | 90 | **440** |
| 15 | 33 | 32 | 33 | 23 | 38 | 36 | 32 | 36 | 34 | 83 | **380** |
| 16 | 52 | 39 | 35 | 13 | 53 | 36 | 34 | 39 | 28 | 75 | **404** |
| 17 | 62 | 39 | 45 | 22 | 44 | 41 | 42 | 32 | 28 | 103 | **458** |
| 18 | 38 | 45 | 43 | 25 | 34 | 47 | 25 | 41 | 24 | 68 | **390** |
| 19 | 45 | 39 | 38 | 17 | 39 | 21 | 24 | 28 | 15 | 51 | **317** |
| 20 | 34 | 39 | 46 | 26 | 45 | 21 | 29 | 28 | 17 | 66 | **351** |
| 21 | 39 | 33 | 45 | 27 | 34 | 34 | 39 | 37 | 22 | 48 | **358** |
| 22 | 44 | 42 | 36 | 32 | 35 | 57 | 58 | 61 | 25 | 56 | **446** |
| 23 | 71 | 37 | 57 | 55 | 54 | 58 | 65 | 62 | 51 | 97 | **607** |
| **24** | **91** | **111** | **77** | **64** | **106** | **85** | **97** | **93** | **77** | **156** | **957** |
| 25 | 53 | 49 | 52 | 55 | 60 | 42 | 44 | 50 | 36 | 70 | **511** |
| 26 | 39 | 35 | 57 | 35 | 43 | 25 | 28 | 25 | 20 | 50 | **357** |
| 27 | 37 | 33 | 23 | 28 | 42 | 19 | 15 | 38 | 22 | 66 | **323** |
| Total | 1360 | 1110 | 1147 | 839 | 1244 | 1104 | 1031 | 1135 | 836 | 1980 | |

Let us look at the results of the first column of measurements (17 FEB 2002) in detail. There were selected 118 pairs of similar histograms related to the interval of 1 hour. For the interval of 2 hours, there were only 65 similar pairs. It is the "neighbour zone effect": the adjacent histograms have the highest chance to be similar. While the interval between histograms increases, the number of similar pairs decreases. However, at 24 hours, this number jumps high (to 91 pair of similar histograms).
Altogether, for the first cell, 697 histogram pairs with the interval of one hour were compared, that included experimental data obtained for almost a month (698 hours, ~ 29 days). The overall number of histogram pairs compared for the first column (as well as for each other column) was over 18000; a total count of selected similar histograms was 1360 (the lower line of the Table 1), i.e., about 7% of number of compared pairs.

Estimation of statistical significance based on hyper-geometric distribution gives the probability of random realization of the "neighbour zone effect" (interval 1 hour) for the first column as **p**<$10^{-9}$. For the interval of 24 hours, it is **p< $10^{-8}$**. The general estimation based on Poisson distribution shows that these two maximums far exceed 99% upper confidence limit.

The total number of pair comparisons of histograms in all ten sets of measurements in the Table 1 was approximately 180,000. The total counts of similar histograms for each time interval are shown in the last "Total" column and at the **Fig.4**. 890 similar pairs correspond to the first interval (p<$10^{-8}$). The count of similar histograms significantly decreases for larger intervals (to 390-430) and then it rises sharply for the interval of 24 hours (to 957, p<$10^{-8}$). The probability that these effects are random is extremely small.

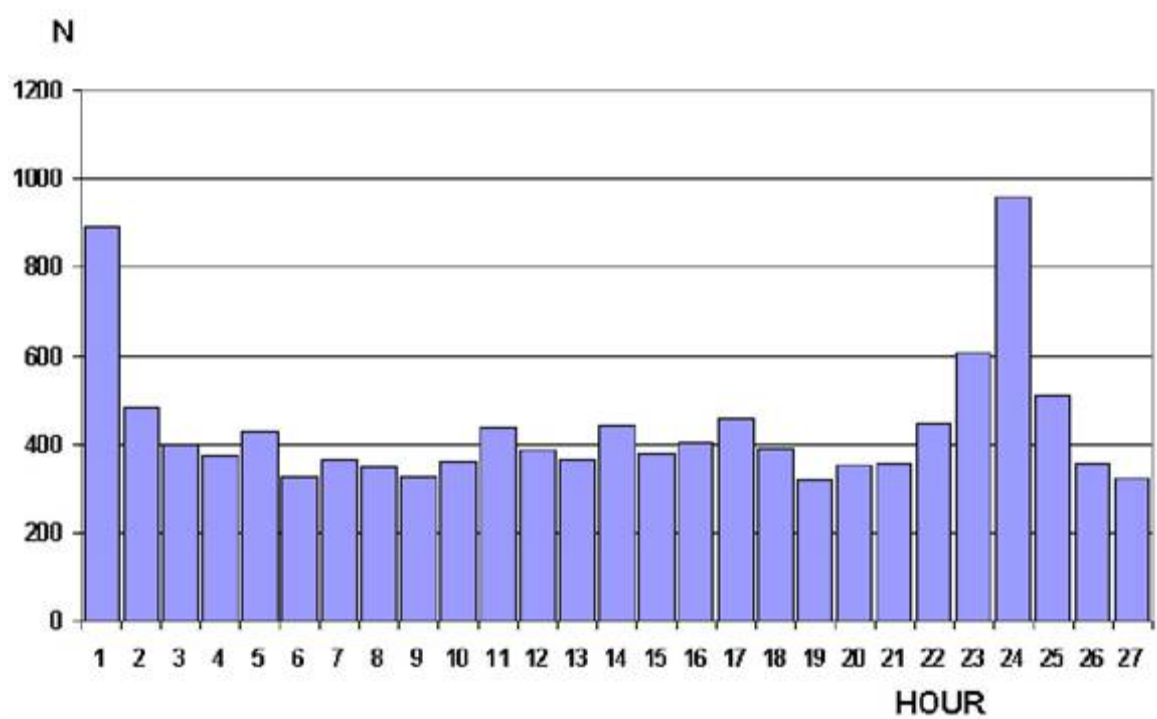

**Fig. 4. Total counts of similar 60-minute histograms for measurements of alpha-activity $^{239}$Pu without collimator. X-axis represents time interval between similar histograms. Y-axis shows a number of similar pairs.**

Table 2 and Fig. 5 represent counts of similar histograms for experiments with the collimator oriented toward the Polar Star. The difference with Table 1 and Fig.4 is obvious: there is neither "neighbour zone effect", nor daily periodicity.

**Table 2.**

**Counts of similar histogram pairs of measurements of $^{239}$Pu alpha-activity with the collimator oriented toward the Polar Star regarding the length of time interval between histograms.**

| Interval (hours) | 17 Feb 2002 | 18 Mar 2002 | 15 Apr 2002 | 14 Jun 2002 | 11 Jul 2002 | 19 Feb 2003 | 19 Feb 2003 a | 18 Mar 2003 | 15 Apr 2003 | 2 Dec 2003 | Total |
|---|---|---|---|---|---|---|---|---|---|---|---|
| 1  | 51 | 46 | 73 | 37 | 55 | 47 | 53 | 56 | 25 | 63 | **506** |
| 2  | 66 | 46 | 55 | 39 | 54 | 24 | 30 | 42 | 20 | 47 | **423** |
| 3  | 74 | 56 | 45 | 39 | 64 | 33 | 37 | 40 | 14 | 35 | **437** |
| 4  | 71 | 44 | 62 | 29 | 47 | 27 | 33 | 26 | 13 | 28 | **380** |
| 5  | 64 | 53 | 64 | 45 | 78 | 41 | 44 | 41 | 16 | 57 | **503** |
| 6  | 55 | 46 | 55 | 35 | 56 | 40 | 41 | 34 | 21 | 45 | **428** |
| 7  | 57 | 47 | 57 | 32 | 68 | 24 | 41 | 31 | 25 | 55 | **437** |
| 8  | 62 | 50 | 66 | 23 | 71 | 25 | 42 | 40 | 15 | 58 | **452** |
| 9  | 67 | 49 | 58 | 25 | 71 | 22 | 34 | 44 | 16 | 47 | **433** |
| 10 | 60 | 50 | 81 | 25 | 54 | 34 | 23 | 46 | 19 | 57 | **449** |
| 11 | 87 | 46 | 72 | 35 | 52 | 29 | 19 | 31 | 16 | 52 | **439** |
| 12 | 76 | 34 | 72 | 35 | 55 | 40 | 31 | 41 | 16 | 39 | **439** |
| 13 | 52 | 50 | 62 | 27 | 48 | 48 | 32 | 41 | 19 | 33 | **412** |

| 14 | 82 | 33 | 53 | 25 | 57 | 46 | 44 | 37 | 32 | 57 | **466** |
| --- | --- | --- | --- | --- | --- | --- | --- | --- | --- | --- | --- |
| 15 | 63 | 40 | 73 | 18 | 51 | 34 | 25 | 36 | 24 | 46 | **410** |
| 16 | 74 | 37 | 75 | 25 | 56 | 37 | 26 | 49 | 18 | 47 | **444** |
| 17 | 95 | 47 | 79 | 34 | 41 | 27 | 22 | 38 | 16 | 43 | **442** |
| 18 | 71 | 51 | 67 | 24 | 61 | 33 | 21 | 42 | 26 | 44 | **440** |
| 19 | 74 | 53 | 66 | 29 | 50 | 21 | 36 | 70 | 17 | 42 | **458** |
| 20 | 67 | 47 | 63 | 34 | 48 | 34 | 38 | 56 | 17 | 53 | **457** |
| 21 | 92 | 44 | 76 | 30 | 53 | 55 | 36 | 41 | 24 | 52 | **503** |
| 22 | 83 | 52 | 53 | 33 | 49 | 45 | 46 | 44 | 24 | 46 | **475** |
| 23 | 75 | 42 | 67 | 32 | 59 | 58 | 49 | 42 | 30 | 62 | **516** |
| **24** | **76** | **47** | **67** | **35** | **62** | **33** | **36** | **56** | **37** | **74** | **523** |
| 25 | 72 | 51 | 72 | 36 | 59 | 37 | 34 | 44 | 40 | 66 | **511** |
| 26 | 86 | 40 | 57 | 28 | 71 | 46 | 36 | 39 | 22 | 56 | **481** |
| Total | 1852 | 1201 | 1690 | 809 | 1490 | 940 | 909 | 1107 | 562 | 1304 | |

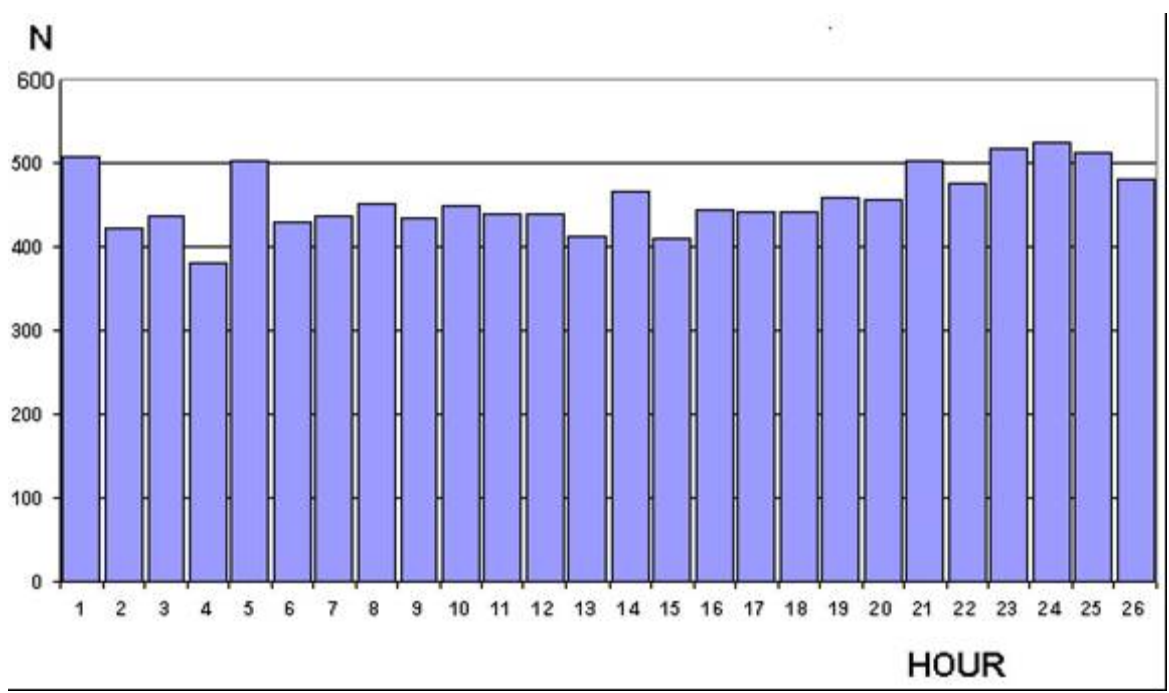

**Fig.5. Counts of similar 60-minute histograms regarding the time interval between them for measurements of alpha-activity $^{239}$Pu with collimator oriented to the Polar Star. Axes as at the Fig. 4.**

As we see at given figures, a chance of repeated appearance of the same shape of histogram is essentially different in measurements with the collimator oriented to the Polar Star and measurements without collimator or when collimator oriented to the West. In the measurements without collimator, or with the collimator oriented to the west, the obvious effect of "neighbor zone" is observed, as well as about-24-hour period. In the measurements with the collimator oriented toward the Polar Star a likelihood of repeated appearance of the same shape of histogram is about the same along time – there is neither "neighbor zone", nor about-daily periodicity. Estimation of probability to obtain such differences in total distributions gives very small values.

About-daily periodicity in appearance of the same shape of histogram in the measurements without collimator, or when collimator oriented to the West, is equal exactly to 1436 minutes, i.e. sidereal day. This period was observed in one-minute histograms constructed for 60 one-second measurements. In these experiments we compared several sets of one-minute histograms. We considered pairs with intervals of 1434 to 1442 minutes between histograms. Each set had 698 histograms. In each experiment, for each column of Table 3, 6282 histogram pairs have been compared, and a count of similar histograms for each time interval has been written into corresponding cell. The average proportion of similar histograms was about 5%. The frequency of appearance of similar histograms at the interval of 1436 minutes is more then 2 times higher then average frequency for other intervals. Estimated probability of random appearance of such effect is less then $10^{-6}$.

**Table 3. Comparison of one-minute histograms of $^{239}$Pu alpha-activity: 24-hour period in histogram shape resemblance is equal 1436 minutes, i.e. sidereal day.**

| | Number of similar histogram pairs | | | | | |
|---|---|---|---|---|---|---|
| Interval (minutes) | 3-4 Jul 2003 | 5-6 Jul 2003 | 12-13 Jul 2003 | 22-23 Jun 2003 | 24-25 Jun 2003 | Total |
| 1434 | 19 | 27 | 31 | 25 | 23 | 125 |
| 1435 | 31 | 31 | 36 | 42 | 35 | 175 |
| **1436** | **97** | **92** | **68** | **95** | **82** | **434** |
| 1437 | 45 | 43 | 35 | 51 | 46 | 220 |
| 1438 | 45 | 32 | 28 | 33 | 33 | 171 |
| 1439 | 29 | 27 | 25 | 26 | 32 | 139 |
| 1440 | 35 | 38 | 21 | 38 | 23 | 155 |
| 1441 | 19 | 15 | 25 | 16 | 20 | 95 |
| 1442 | 9 | 13 | 11 | 19 | 18 | 70 |
| Total | **329** | **318** | **280** | **345** | **312** | **1584** |

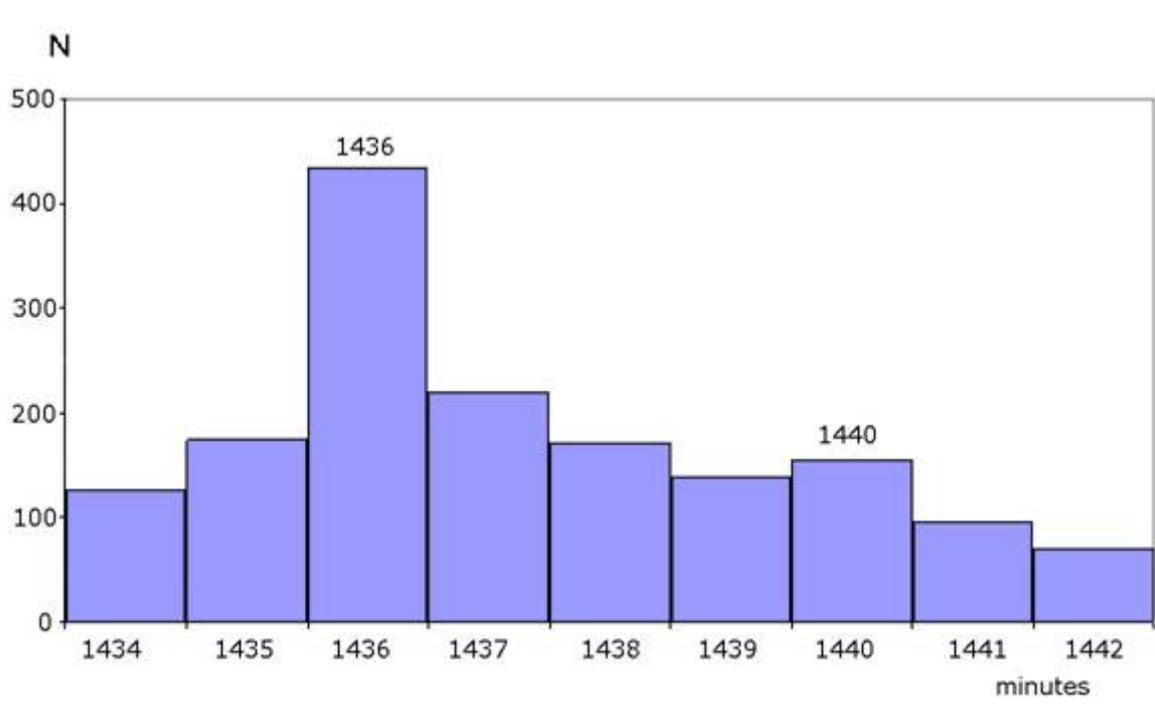

**Fig. 6.** Count of similar histogram pairs for measurements of $^{239}$Pu alpha-activity with collimator oriented to the West. About-daily period of appearance of the same histogram shape is equal to 1436 minutes. X-axis is intervals between histograms in minutes. Y-axis is total number of similar histogram pairs.

## Conclusions

The results exclude any trivial explanation. False effects are excluded not only by strict following to the principle of "equality of other conditions"( *ceteris paribus* principle) – the only difference between "case" and "control" is direction of alpha particles registered in experiments. It is absolutely impossible to affect the radioactive decay itself. It is paradoxical that in experiments conducted on 54º North latitude with collimator oriented toward the Polar Star the shapes of histograms vary in time by the same way as histograms received for measurements near the North Pole (Shnoll S.E., 2003).

It may indicate that probability of flow of alpha particles is not the same for different directions, and variations of this probability in time depend on changes in the surrounding time-space continuum. The presence of clear periodicity in these changes with the sidereal period of 1436 minutes (Shnoll S.E., 2001), the higher frequency of the appearance of similar histograms at the same local (longitudinal) time, the disappearance of daily periodicity near the North Polar, and, finally, presented above results of the experiments with collimators, - all these facts suggest the dependence of the observed effects on the state of the celestial sphere. The registration of such dependence is not sufficient for declaring a hypothesis, which might explain its "mechanism". However, in this article our task is only to register the significance of these astonishing phenomena.

**ACKNOWLEDGEMENTS**

We are indebted to M.N. Kondrashova and L.A. Blumenfeld who inspired, encouraged, and discussed this work. Special thanks are to Y.M. Popov who built the mechanical part of detectors used in this study. We are grateful to our colleagues, T.A. Zenchenko, A.A. Konradov, and S.S. Zhirkov for collaboration and assistance; and to V. K. Lyapidevskii, B.M. Vladimirsky, V.A.Tverdislov, A.P.Levich, B.V. Komberg, A.S. Kondrashov, F.A. Kondrashov, D.P. Kharakoz, F.I. Ataullakhanov, V.N. Morozov, I.M. Dmitrievsky, Yu.S.Vladimirov, D.S.Chernavsky and B.U. Rodionov for valuable discussions. Financial support obtained from V.P. Tikhonov and T. Peterson and their interest to this work are gratefully acknowledged.

Correspondence and requests for materials should be addressed to S.E.S. (e-mail: shnoll@iteb.ru).